# A James Webb Space Telescope NIRCam Deep Field Simplified Simulation Using a Geometric-Focused Ensemble Approach

**Matthew Sailer[1]**

[1] Department of Space Studies, American Military University, Charles Town, West Virginia, United States

E-mail: matthew.sailer@mycampus.apus.edu; mwsailer12@gmail.com

## Abstract

Recent studies predict the characteristics of JWST's deep field image using a deterministic approach based on recent observational measurements with corresponding ranges of uncertainty. This study presents a simplified geometric-focused deep field simulation using an ensemble approach to demonstrate the high variability in results due to the uncertainty ranges of observational measurements. Two sets of ensemble simulations were conducted: a parameter sensitivity ensemble, and a 90-member full ensemble each calculating the percentage of the image occupied by galaxies. The estimated number of unseen galaxies in the HUDF was found to provide the largest variability of results. The Apparent Galaxy Wall (AGW) effect is introduced and defined as $\geq 50\%$ area of a deep field image occupied by galaxies. A one-way one-sample t-test was conducted on the 90-member ensemble, concluding the JWST is not likely to observe the AGW effect with an estimated galaxy coverage percentage of 47.07% ± 31.85 but does not rule out the effect as a possibility. A discussion is included on the potential impacts of the AGW effect being observed and its potential to form a pseudo-cosmological horizon that may inhibit the effectiveness of future observatories.

## 1. Introduction

After years of delays, the James Webb Space Telescope (JWST) is expected to launch before the end of 2021, and countless scientists, researchers, and space enthusiasts eagerly look forward to the release of its first images (Fisher 2021). The four primary research areas of JWST are defined as (i) The End of the Dark Ages: The First Light and Reionization, (ii) The Assembly of Galaxies, (iii) The Birth of Stars and Protoplanetary Systems, and (iv) The Planetary Systems and the Origins of Life (Gardner et al. 2006). The JWST has been described as the successor to the Hubble Space Telescope (HST) and designed to observe in longer wavelengths to probe deeper into the observable universe (Gardner et al. 2006; NASA 2020a). The primary research areas and status as the successor to HST indicate the JWST is likely to produce some of the furthest-reaching deep field images ever captured with distances expected to reach redshifts as large as z = 15 with the NIRCam instrument (Gardner et al. 2006).

One of the JWST mission objectives is to study the Hubble Ultra Deep Field (HUDF) and The Great Observatories Origins Deep Survey (GOODS) region in greater depth with its highly sensitive infrared imager Mid-InfraRed Instrument (MIRI) and Near-Infrared Camera (NIRCam) respectively (Villard 2017).

Given the NIRCam specifications, a numerical simulation can shed light on the expected JWST deep field images before release. A novel simulation was developed of the JWST deep field images using a heavy emphasis on general characteristics of the telescope and recent observational measurements of the universe. The python code written for this study can be downloaded from the Astrophysical Source Code Library (Sailer 2021).

*1.1 Necessity for Study*

Although several studies predict the characteristics of the JWST deep field images from accepted methods, it is impossible to determine a method's accuracy until the images are released. Despite the completed studies and simulations, alternate and novel methods of simulation provide added benefit to existing simulations by broadening the range of potential outcomes. This study aims to simulate the JWST deep field images by focusing on the typical galaxy's apparent two-dimensional geometry through high-z interpolation of recent observation-based studies and using an ensemble method through the perturbation of the initial conditions. The ensemble approach demonstrates the high variability in possible solutions not reflected in other similar studies. The strength of this study is found through the ensemble method used to capture this variabililiy in a quantitative manner.

*1.2 The Apparent Galaxy Wall (AGW) Effect*

One of the unique predictions resulting from this simulation is an apparent wall of galaxies resulting from a large galaxy number density and increased apparent angular diameters of high-z galaxies due to the expansion of the universe as predicted by general relativity. The confusion limit is a well known limit to many long-wavelength observatories where poor spatial resolution and a high density of high-z sources blends together into the Extragalactic Backround Radiation (Vaisanen et al 2001; Windhorst et al. 2008; Cooray 2016). Separate from the confusion limit, an additional limit is shown to be possible under the right conditions through a perturbations of the galaxy number density based on the number of galaxies theorized to remain unseen in the HST due to detector limitations. This theoretical visually obstructing barrier may even act as a pseudo-cosmological horizon if it occurs in nature. The AGW effect is defined in this work as occurring when $\geq$ 50% of a given area is visually obstructed by galaxies due to galaxy number density saturation. Despite the AGW effect's demonstration in some of the ensemble members, this simulation does not predict a large liklihood of the JWST observing the AGW effect in its deep field images.

*1.3 Scope of Research*

The geometry focus of this research refers to the study of the total area of a deep field image occupied by galaxies through calculations applying the basic equations that govern space-time geometry and the expanding universe. The scope of this research focuses on the perturbations of the initial conditions and thus makes some simplifications due to the large number of runs required for an ensemble simulation. Because of this, gravitational lensing events and clustering events are not considered. Galaxies are represented as ellipses with dimensions defined based on effective radius. The radial light profile PSFs are approximated by applying a Gaussian filter to the simulated ellipses. A general initial mass function (IMF) is assumed and tailored to the characteristics of 6 different types of galaxies: irregular, elliptical, spiral, lenticular, dwarf spheroidal, and dwarf elliptical. Finally, this research simulates a random spatial distribution of galaxy placement with a number density dependent on the redshift.

## 2. Background

*2.1 Previous Deep Field Images*

The HUDF image is arguably the most famous deep field image in astronomy with an estimated 10,000 galaxies in one image and often serves as a comparison for deep field images and early universe analysis (Beckwith et al. 2006; Inami et al. 2017; NASA 2018). Other notable deep field images include the Extreme Deep Field, the Spitzer Cosmic Assembly Near-Infrared Deep Extragalactic Survey (S-CANDELS), and the GOODS Re-ionization Era wide-Area Treasury from Spitzer (GREAT) survey to name a few (Illingworth et al. 2013; Ashby et al. 2015; De Barros et al. 2019). Because of the substantial analysis performed on the HUDF, it was chosen as the calibration source for this simulation. The JWST is expected to see further than ever before and generally outperform previous deep field images due to its large primary mirror, advanced detectors, near-IR wavelengths detection, and high planned exposure time (Gardner et al. 2006; NASA 2020a).

*2.2 Previous Simulations*

There have been several simulations of the upcoming JWST deep field images generated in recent years that do not necessarily depict the AGW effect occurring (ESA 2002; Gardner et al. 2006; ESO 2011; Kauffmann et al. 2020). Additionally, some catalogs have been created and surveys performed that predict the characteristics of distant galaxies expected to be observed by the JWST or relating effective radius to redshift. These catalogs and surveys are based on collected data from the JAdes extraGalactic Ultradeep Artificial Realizations (JAGUAR) catalog, the FourStar Galaxy Evolution Survey (ZFOURGE), and the Multi-Unit Spectroscopic Explorer survey (MUSE) to name a few (Spitler 2012; Herenz et al. 2017; Williams et al. 2018). No research has predicted galaxy number density saturation from the results of these surveys and catalogs at the time of writing. While these catalogs are built from recent observations and offer valuable insight, they are limited by the constraints of the detectors used to build the catalogs. The number densities of these catalogs decrease as redshift increases making it difficult to obtain an accurate number density count to include the unseen galaxies in their respective distance ranges. A different approach is used through the integration of comoving volume and the galaxy merger rate after calibrating the number densities through the replication of the HUDF image.

Apart from the JWST predictions based on catalogs and surveys, several visual simulations are presented on several organizational websites including the European Space Agency and the European Southern Observatory (ESA 2002; ESO 2011). While mock catalogs can provide simulated characteristics of galaxies in the early universe, it is the visual simulations of these mock catalogs that provide a clear representation of what may be seen in the upcoming JWST deep field images. These visual simulations appear to be based on some of these surveys, but little supporting documentation or publications on how these simulations were calculated were found online apart from ESA (2002).

Additionally, a commonly used programme called SkyMaker has been used as an accepted simulation programme for deep field images (Bertin 2008). The resulting simulations assume a universe of flat geometry with $\Omega_m = 0.3$, $\Omega_\Lambda = 0.7$.

While there are many simulations of the early universe, not all of them focus on the upcoming JWST deep field images. Only simulations that focus on the JWST will be discussed in this project, and other simulations which are not directly aimed at predicting the JWST like the EAGLE Project and the IllustrisTNG Project were not specifically considered (Schaye et al. 2014; Crain et al. 2015; Marinacci et al. 2018).

The most recent simulation with substantial documentation was published by Kauffmann et al. (2020) building off the mock catalog and visual simulation by Williams et al. (2018). Kauffmann et al. (2020) produced a deep field simulation using the JAGUAR catalog and assumed $\Omega_m = 0.3$, $\Omega_\Lambda = 0.7$, $H_0 = 70 \frac{km}{s \cdot Mpcs}$, a resolution of $0.031''$/pixel, a ~29 AB mag (5σ), and an area of 4.5 arcmin$^2$.

*2.3 The Limitations of Current Simulations*

While these previous simulations excel in many areas and produced great visuals, this study does not use their approach for the following reasons. First, the simulations with included documentation use catalogs like the JAGUAR catalog for mock sources which imply the JWST is likely to observe only 10s of galaxies at z > 10 (Williams et al. 2017; Kauffmann et al. 2020). This estimate appears to contradict expected numbers when considering the HUDF. The current record for the highest redshift measured in the HUDF region is z ~ 11 (Oesch et al. 2016; Jiang et al. 2020; Schauer, Drory & Bromm 2020). When considering the HUDF MUSE survey successfully measured the redshifts of ~15% of the galaxies in the HUDF, and considering closer galaxies are generally easier to obtain a redshift value than galaxies at high-redshifts due to a generally greater photon flux, it can be reasonably concluded that a significant number of galaxies imaged in the HUDF have redshifts of z > 10 but are too faint to obtain redshifts from their spectra (Inami et al. 2017). Similarly, when considering the JWST is expected to see objects 100 times fainter than the HST, and that early galaxies are theorized to contain hot and bright massive stars, it can be subjectively concluded that tens of galaxies with redshifts of z > 10 is a substantial underestimate (Gardner et al. 2006; Larson & Bromm 2009; Benson 2010; Hashimoto et al. 2018; Inami et al. 2017). Further evidence for this can be found when considering the deep field image captured by the Spitzer Space Telescope depicting hints of the AGW effect with the appearance of a high galaxy number of density at large redshifts (NASA 2019). This appearance, however, is due to the confusion limit due to observation in long wavelengths where the PSF of galaxies up to z ~ 3 - 3.5 is overlapping due to the wavelengths detected by Spitzer in the 3-10 μm range (Vaisanen et al 2001).

Second, several articles theorize that there may exist 2 - 10 times as many galaxies as seen in the HUDF's field of view that were not detected due to the limitations of the HST (Hille 2016; Conselice et al. 2016; Lauer et al. 2021). Because of the difficulty of modeling an uncertain galaxy number density, galaxy catalogs alone may not offer the best foundation on which to build an accurate simulation of the early universe.

Third, the functions used to interpolate backward in time heavily rely on the observations of previous observatories which Williams et al. (2018) admit are limited by spatial coverage, sensitivity, and atmospheric effects of the HST, SST, and ground-based observatories, respectively. Relying on observations alone may not yield the most accurate galaxy number density estimates for JWST simulations. Using a general fundamental approach using Comoving Volume may offer a more accurate representation of galaxy numbers.

Finally, the major constants that describe the universe are assumed as constant values for these simulations with no perturbations to explore the effects of changes to these constants. Since many of these measurements contain significant uncertainty ranges, the values of the initial conditions affect a simulation's result. The initial conditions are perturbed to generate the full range of possible results to demonstrate the importance of an ensemble approach.

Because of these considerations, a novel simulation was constructed in python assuming the general specifications of the JWST as outlined by Gardner et al. (2006) focuses on the galaxy coverage percentage in the deep field image using cosmological equations calibrated with the HUDF.

## 3. Method

An original code was written in python to produce a geometric-focused JWST deep field simulation that can be run with an ensemble method. A randomly generated deep field image is produced from calculations based on general equations derived from recent observations. The resulting simulation represents galaxies as randomly oriented ellipses with random angular orientation, random spatial placement, systematically generated galaxy types, and random edge-on to top-down orientation for galaxies with elliptical geometries. In addition to a deep field image, the code produces a calculation of the percentage of background space obstructed by foreground galaxies and a few other supporting figures that will be discussed in later sections. The percentage of background space obstructed by foreground galaxies by area will be referred to as "galaxy coverage percentage" throughout the rest of this paper. The code's effectiveness was tested by inputting initial conditions of the HST and successfully producing a similar-looking image to the HUDF to demonstrate the effectiveness of the methods used in this simulation. There are ten perturbed initial conditions with each perturbation run for nine NIRCam filters. Each perturbation was simulated to produce 90 unique results for a 90-member ensemble. The galaxy coverage percentages associated with each unique result were averaged and compared to the AGW definition of $\geq 50\%$ using a one-way one-sample t-test.

### 3.1 Estimates and Initial Conditions

Several general estimates were made to create a simple, yet effective, simulation. The generally accepted ΛCDM model was assumed for this simulation due to successful observational tests to include CMB, supernovae, galaxy clustering, and weak lensing observations (Conley et al. 2011; Zehavi et al. 2011; Suzuki et al. 2012; Bennett et al. 2013; Heymans et al. 2013; Planck Collaboration et al. 2016; Hildebrandt et al. 2017; Wang et al. 2018). Even though the ΛCDM model was assumed, no specific values of constants were assumed, but rather, the constants describing the ΛCDM model were perturbed through their ranges of uncertainty from recent measurements.

#### 3.1.1 JWST Characteristics

The JWST's NIRCam characteristics can be found in its user documentation and summary publications. NIRCam was designed to have a sensitivity reaching an AB magnitude of 31 from 100 - 200 hours of exposure per filter resulting in the ability to detect sources 100 times fainter than the HST and reaching a redshift of z > 15 (Gardner et al. 2006). Additionally, STSci (2016) and Villard (2017) provides an overview of the JWST NIRCam instrument from the James Webb Space Telescope User Documentation and specifies it will have a spatial resolution as low as 0.031 arcsec per pixel for the NIRCam. This high spatial resolution and sensitivity are in part due to a significantly larger segmented primary mirror allowing for 6.25 times more light-collecting area with highly sensitive sensors (NASA 2020a). This study chose to simulate deep field images using an exposure of 10 ks in the following filters: F090W, F115W, F150W, F200W, F277W, F335M, F356W, F410M, and F444W to match previous simulations and intended deep field image specifications (Gardner et al. 2006; Kauffmann et al. 2020).

### 3.1.2 General cosmological considerations

The early universe is difficult to thoroughly study through observations alone due to our observational limitations (Larson & Bromm 2009). Because of this, the following generalizations were made given the respective supporting publications. (i) Galaxy formation is assumed to have occurred between $20 < z < 50$ as described by Benson (2010). This code assumes no galaxy formation occurred outside of this redshift range when estimating galaxy number densities. (ii) The first stars are estimated to have a large mass and high luminosity (Gardner et al. 2006) resulting in the first galaxies also having a high luminosity (Benson 2010). Evidence for this can be found in several studies like the Butcher-Oemler effect (Butcher & Oemler 1984), the brighter-than expected high-z galaxies captured by SST (De Barros et al. 2019), higher than expected galaxy maturity found in the ALPINE-ALMA [CII] survey (Le Fèvre et al. 2020), and the recent announcement from NASA (2020b). (iii) This study sets as an initial condition the age of the universe when star formation began and assumes the value to be between 100 million and 250 million years after the big bang (Larson & Bromm 2009). (iv) The IMFs of simulated galaxies are generated with a dependency on this value and galaxy type.

### 3.1.3 Previous Measurements of $H_o$, $\Omega_m$, $\Omega_\Lambda$, $\Omega_k$, and the Dark Energy Equation of State

Numerous studies have calculated estimates for $H_0$, $\Omega_m$, $\Omega_\Lambda$, $\Omega_k$, and the Dark Energy Equation of State. Many recent measurements were gathered to determine a range of possible values using their uncertainties.

The Hubble Constant, $H_0$, has been measured to be: $70.4^{+1.3}_{-1.4} \frac{km}{s \cdot Mpc}$, $73 \pm 2$ (random) $\pm 4$ (systematic) $\frac{km}{s \cdot Mpc}$, $68.5 \pm 3.5 \frac{km}{s \cdot Mpc}$, $68.3^{+2.7}_{-2.6} \frac{km}{s \cdot Mpc}$ in a spatially flat $\Lambda$CDM model and $68.4^{+2.9}_{-3.3} \frac{km}{s \cdot Mpc}$ in a non-spatially flat $\Lambda$CDM model, $73.52 \pm 1.62 \frac{km}{s \cdot Mpc}$, $72.56 \pm 1.5 \frac{km}{s \cdot Mpc}$, $69.8 \pm 0.8$ ($\pm 1.1\%$ stat) $\pm 1.7$ ($\pm 2.4\%$ sys) $\frac{km}{s \cdot Mpc}$, $82^{+8.4}_{-8.3} \frac{km}{s \cdot Mpc}$, $68.20 \pm 0.81 \frac{km}{s \cdot Mpc}$, and $67.4 \pm 0.5 \frac{km}{s \cdot Mpc}$ (Benson 2010; Freedman & Madore 2010; Verde, Protopapas & Jimenez 2014; Chen, Kumar & Ratra 2017; Riess et al. 2018; Capozziello, Ruchika & Sen 2019; Freedman et al. 2019; Jee et al. 2019; Alam et al. 2020; Planck Collaboration et al. 2020). In recent years, estimates for $H_0$ appear to diverge depending on the method used to estimate it. These 2 values are $\sim 67 \frac{km}{s \cdot Mpc}$ and $\sim 73 \frac{km}{s \cdot Mpc}$ (Castelvecchi 2020). The mass density of the universe, $\Omega_m$, has been measured to be: $0.2726 \pm 0.014$, $0.295 \pm 0.034$ (stat+sys), $0.32 \pm 0.05$, and $0.315 \pm 0.007$ (Benson 2010; Betoule et al. 2014; Verde et al. 2014; Planck Collaboration et al. 2020). The dark energy density of the universe, $\Omega_\Lambda$, has been measured to be: $0.728^{+0.015}_{-0.016}$ and $0.6847 \pm 0.0073$ (Benson 2010; Planck Collaboration et al. 2020). The total energy density of the universe, $\Omega_k$, has been measured to be: $-0.02 \pm 0.24$, $-0.0023 \pm 0.0022$, and $0.001 \pm 0.002$ (Wei & Wu 2017; Alam et al. 2020; Planck Collaboration et al. 2020). The dark energy equation of state, w, has been measured to be: $-1$, $-1.027 \pm 0.055$, $-1$, $-0.912 \pm 0.081$, and $-1.03 \pm 0.03$ (Huterer & Cooray 2005; Betoule et al. 2014; Moews et al. 2019; Alam et al. 2020; Planck Collaboration et al. 2020). This extensive review was necessary to determine the perturbation range used in the simulation. Each measurement was averaged together to obtain a mean estimate which served as a baseline value for the simulation. Only $H_0$ estimates based on high-z observations are considered in this baseline's average as focusing on measurements within a small distance from the Milky Way can yield inaccurate results (Alam et al., 2020). The maximum and minimum values of the ranges of uncertainties were identified and used as the perturbation ranges in the ensemble runs.

### 3.1.4 Proper Length Estimate

The proper length of a galaxy is strongly dependent on galaxy type. There are some studies which estimate the average proper length of galaxies at high redshifts, but there is a wide range of results.

Ono et al. (2013) estimate the effective radii of galaxies between 7 < z < 12 are 0.3-0.4 kpc and evolve proportionally to $(1+z)^{-1.28\pm0.13}$ from observations of the HUDF. Allen et al. (2017) estimate an effective radius evolution seen in Equation 1 from stellar-mass $\log(\frac{M_*}{M_\odot}) > 10$ from redshifts of 1 < z < 7.2 using samples from star-forming galaxies in the FourStar Galaxy Evolution Survey (ZFOURGE).

$$r_e = 7.07(1+z)^{-0.89\pm0.01} \text{ kpcs} \quad (1)$$

Williams et al. (2018) offer a proper length estimate for high-z galaxies; however, the estimate is split into several different types of galaxies. The effective radii of high-z dwarf galaxies are more difficult to generalize due to observational limitations and dwarf galaxy brightness. This study selects an estimated proper length for each galaxy by assuming Allen et al.'s (2017) for spiral, elliptical, lenticula, and irregular galaxies for all values of z > 1 while assuming typical dwarf galaxy proper lengths from their observed ranges of length. A more detailed systematic approach to proper length estimation and additional characteristics is discussed in section 3.3.4.

### 3.1.5 Galaxy Merger Rate

Because the comoving volume is used to estimate the galaxy number density, the galaxy merger rate is a necessary consideration. The galaxy merger rate is difficult to determine since the large-scale motion of the universe is minuscule compared to the average human lifetime and limitations of current observatories, but estimates can still be made from observations (Conselice et al. 2016). The universe has evolved into a state of higher structure since the Big Bang due to the gravitational attraction of galaxies into clusters and superclusters, and many of the galaxies today, including the Milky Way, are believed to have undergone several merger events in the past (Lotz et al. 2011). This adds another level of complexity in determining the galaxy merger rate (Postman 2001). Conselice et al. (2016) estimate the total number of galaxies in the universe declines with time proportionally to $\sim\frac{1}{t}$ where t is the age of the universe in Gyrs from observations out to z = 8. For simplicity, the estimated merger rate is assumed for all values of z.

### 3.1.6 HUDF Unseen Galaxies

A deep field image cannot capture all existing galaxies due to the variation of galaxy sizes and luminosities. Recent studies estimate there exists between 2 - 10 times as many galaxies as seen in the HUDF with redshifts 0 < z < 8 which are unable to be detected by the HST (Hille 2016; Conselice et al. 2016; Lauer et al. 2021). The most recent estimate is from the New Horizons' Cosmic Optical Background (COB) that estimates only two times as many galaxies exist (Lauer et al. 2021). This study will assume the JWST is likely to see many of these unseen galaxies in the smaller redshift bins due to its ability to see objects 100 times fainter than the HST (Gardner et al. 2006), the unexpectedly high luminosity of galaxies in the early universe (Butcher & Oemler 1984; De Barros et al. 2019), the higher than expected galaxy maturity in the early universe (Le Fèvre et al. 2020), and recent measurements of the age at which stars first formed (NASA 2020b).

### 3.1.7 Anisotropy of Dark Energy

Recent studies have suggested the nature of dark energy may be anisotropic. (Kolatt & Lahav 2001; Schwarz & Weinhorst 2007; Kalus et al. 2013; Campanelli et al. 2011; Mariano & Perivolaropoulos 2012; Cai et al. 2013; Cooke & Lynden-Bell 2010; Antoniou & Perivolaropoulos 2010; Heneka et al. 2014; Zhao et al. 2013; Yang et al. 2013; Javanmardi et al. 2015; Lin et al. 2015; Chen & Chen 2019; Colin et al. 2019; Migkas et al. 2020; Porter & Watzke 2020). A novel consequence of the anisotropy of dark energy was considered. The ratio of anisotropy being perturbed as an initial condition in the ensemble simulation. This study assumed the anisotropy of dark energy is consistent in nature by generalizing the rate of expansion as proportional to a direction-dependent Hubble constant for simplicity.

### 3.2 General Approach

The following ten initial conditions were perturbed: (i) the Hubble constant ($H_0$), (ii) the mass density ($\Omega_m$), (iii) the dark energy density ($\Omega_\Lambda$), (iv) the Dark energy equation of state parameter (w), (v) the number of unseen galaxies by the HST (n_Unseen), (vi) the increase in the percentage of each galaxy the JWST will be able to see from an increase in sensitivity compared to HST (r_Increase), (vii) the ratio of the Hubble constant in the y-axis to the Hubble constant in the x-axis due to the anisotropy of dark energy (DE_Ratio), (viii) the maximum redshift expected to be measured by the JWST (zMax), (ix) the universe's age at which star formation first began (ageOfStarBirth), and (x) the observed filter used by the JWST NIRCam (jwst_Filter). A parameter-sensitivity ensemble was run by choosing a baseline set of initial conditions and perturbing each initial condition individually through its corresponding observation-based uncertainty range before calculating the corresponding change to the galaxy coverage percentage. The standard deviation of each initial condition's perturbation result was used as a measure of its sensitivity with the most sensitive parameters having the highest standard deviation. A full ensemble was run by perturbing all initial conditions simultaneously through their observation-based uncertainty ranges and calculating the corresponding galaxy coverage percentages. A one-way one-sample t-test was used to analyse the resulting 90 galaxy coverage percentages compared with the AGW effect's definition of $\geq 50\%$ coverage.

This work aimed to take advantage of the powerful ensemble approach to computation. Since computing has been described as the third pillar of science, the method of computation was a vital consideration for an effective simulation, so an ensemble approach was chosen (Skuse 2019; Winsberg 2019). The ensemble approach tends to be underused in science but is extensively used in numerical weather prediction with effective results (Cahir n.d.; NOAA n.d.; JMA n.d.; Smagorjnsky 1983; Gneiting & Raftery 2005; Environment Canada 2013; Andersson 2014; Slater, Villarini & Bradley 2016; Kikuchi et al. 2017). Additionally, the ensemble approach has been effectively used in multiple studies including water resource studies, space weather, and gene expression to name a few demonstrating its effectiveness as a computational method (Knipp 2016; Mays et al. 2015; Quilty, Adamowski & Boucher 2019; Moyano et al. 2021)

### 3.2.1 Parameter-Sensitivity Ensemble

A preliminary ensemble simulation was run to test the sensitivity of each initial condition. A baseline set of initial conditions was defined as a control. The baseline was chosen to be $H_0 = 71.4$, $\Omega_m = 0.301$, $\Omega_\Lambda = 0.706$, $w = -0.994$, n_Unseen = 6, r_Increase = 130%, DE_Ratio = 0.925, zMax = 16, ageOfStarBirth = $1.75e8$, and jwst_Filter = 'F277W.' Each initial condition was run through a range of 10 values while the rest of the initial conditions were held constant.

The resulting galaxy coverage percentages were averaged together for each initial condition parameter and the standard deviation was calculated. The parameters were ranked from most sensitive to least sensitive based on their standard deviations.

### 3.2.2 Ensemble Simulation

A full ensemble simulation was run perturbing all initial conditions simultaneously to calculate the range of possible resulting galaxy coverage percentages. Since the initial condition ranges are defined by the uncertainties of recent measurements, this ensemble represents the full range of possible JWST deep field outcomes. The resulting galaxy coverage percentages were saved in a list, and the mean and standard deviation of this list was calculated.

### 3.2.3 T-test Analysis

A one-way one-sample t-test was used to compare the resulting mean and standard deviation from the ensemble simulation with the AGW definition of $\geq 50\%$ galaxy coverage percentage. The statistical significance was calculated for this prediction. A one-way one-sample t-test was used due to its simplicity, robustness, and ease of calculation (Jackson 2016; UCLA 2016; Flom 2018; UCLA 2019). For this test, with a mean galaxy coverage percentage of $< 50\%$, it could be concluded that the AGW effect is not likely to be observed by the JWST. Additionally, if the final galaxy coverage percentage mean is $\geq 50\%$, then the one-tailed one-sample t-test can be conducted to test for 95% confidence. For this t-test, the alternative hypothesis ($H_a$) was defined as: The JWST will observe a galaxy coverage of $\geq 50\%$, and the null hypothesis ($H_0$) was defined as: The JWST will observe a galaxy coverage of $< 50\%$.

## 3.3 Calculations

The following section discusses the calculations made to create the JWST Deep Field simulation. The following modules were necessary for this code to operate: numpy, quad from scipy.integrate, math, matplotlib, Ellipse from matplotlib.patch, cv2, scipy.ndimage, and Bbox from matplotlib.transforms.

### 3.3.1 Control Panel

The control panel defines the values of the initial conditions to include image resolution, image dimensions, the speed of light, Planck's constant,

Coulomb's constant, exposure time, Extragalactic Background Light (EBL) intensity, primary mirror area, the maximum redshift detected in the HUDF, all perturbed initial conditions used for the ensemble simulation, and an arbitrary length and redshift used to test and plot the angular-diameter-redshift relation.

*3.3.2   Preliminary Calculations*

The following preliminary calculations are made before any simulation is attempted. First $\Omega_k$ is calculated according to Equation 2 (Weinberg 1972; Weedman 1988; Sahni & Starobinsky 2000; Carroll & Ostlie 2018; Peebles 2020). Second, *m* is calculated according to the dark energy equation of state in Equation 3 as discussed by Moews et al. (2019) where *w* is the dark energy equation of state parameter and assumed to be constant. Third, the universe's age (*t*) is calculated according to Equations 4 and 5 (Hogg 2000; Sahni & Starobinsky 2000; Croton 2013; Chen et al. 2017; Balakrishna Subramani et al. 2019).

$$\Omega_k = \Omega_\Lambda + \Omega_m \quad (2)$$

$$m = 3(1 + w) \quad (3)$$

$$E(z) = \sqrt{(1-\Omega_k)(1+z)^2 + \Omega_m(1+z)^3 + \Omega_\Lambda(1+z)^m} \quad (4)$$

$$t = \frac{1}{H_0} \int_0^\infty \frac{dz'}{(1+z')E(z')} \quad (5)$$

Fourth, the galaxy number density is estimated using comoving volume (Hogg, 2000) and Conselice et al.'s (2016) merger rate estimate. A list of redshifts is calculated in intervals of 0.5 Glyrs out to the redshift defined in the zMax initial condition. A galaxy number density is calculated for each redshift using the following method.

The integral seen in Equation 6 is constructed to calculate the number of galaxies between 2 redshifts where "Galaxy Number" is the number of galaxies in an arbitrary field of view between redshifts of $z_1$ and $z_2$, $q$ is an integration constant, $(1+z')$ is the galaxy merger rate estimated by Conselice et al. (2016), and $V_c(z')$ is the comoving volume as described by Hogg (2000) and defined in Equations 7-10.

$$\text{Galaxy Number} = \int_{z_1}^{z_2} q(1+z') V_c(z') dz' \quad (6)$$

$$V_c V_c = \begin{cases} \frac{4\pi}{2\Omega_k} D_H^3 \left[ \frac{D_M}{D_H} \sqrt{1 + \Omega_k \frac{D_M^2}{D_H^2}} - \frac{1}{\sqrt{|\Omega_k|}} \text{arcsinh}\left(\sqrt{|\Omega_k|} \frac{D_M}{D_H}\right) \right] & \text{for } \Omega_k > 1 \\ \frac{4\pi}{3} D_M^3 & \text{for } \Omega_k = 1 \\ \frac{4\pi}{2\Omega_k} D_H^3 \left[ \frac{D_M}{D_H} \sqrt{1 + \Omega_k \frac{D_M^2}{D_H^2}} - \frac{1}{\sqrt{|\Omega_k|}} \text{arcsin}\left(\sqrt{|\Omega_k|} \frac{D_M}{D_H}\right) \right] & \text{for } \Omega_k < 1 \end{cases} \quad (7)$$

$$D_M = \begin{cases} D_H \frac{1}{\sqrt{|\Omega_k|}} \sinh\left(\sqrt{|\Omega_k|} \frac{D_C}{D_H}\right) & \text{for } \Omega_k > 1 \\ D_C & \text{for } \Omega_k = 1 \\ D_H \frac{1}{\sqrt{|\Omega_k|}} \sin\left(\sqrt{|\Omega_k|} \frac{D_C}{D_H}\right) & \text{for } \Omega_k < 1 \end{cases} \quad (8)$$

$$D_C = D_H \int_0^{z'} \frac{dz''}{E(z'')} \quad (9)$$

$$D_H = \frac{c}{H_0} \quad (10)$$

By setting "Galaxy Number" = 10,000 according to the HUDF estimate (Villard 2017; NASA 2018), $z_1$ = 0, and $z_2$ = 11 according to the maximum redshift calculated in the HUDF (Oesch et al. 2016; Jiang et al. 2020; Schauer et al. 2020), a value of *q* is calculated. Equation 6 is then used to calculate a galaxy number density between each previously chosen redshift. These galaxy number densities are saved in a separate list and associated with the corresponding $z_2$ used in the calculation thus producing a slight underestimate for galaxy number densities. Each galaxy number density is then divided by the total area of the HUDF to calculate the density per arcmin$^2$. If the resulting galaxy number densities are less than 1 galaxy per arcmin$^2$, then the value is assigned 1. This estimation is made for the following reasons. (i) In future calculations, if a galaxy number density is less than 1-galaxy per arcmin$^2$, the simulation tends to generate no galaxies due to whole number rounding. (ii) Due to the inherent structure of the universe, there is a higher-than-average local density of galaxies. This approach inherently assumes a uniform galaxy distribution that is not observed in nature. By rounding the low-redshift galaxy number densities up to 1-galaxy per arcmin$^2$, a higher local galaxy density is simulated with little effect to the results since few low-redshift galaxies appear in deep field images compared to high-z galaxies.

Fifth, after a galaxy number density is calculated for each redshift, a proper length is calculated in a separate list for each redshift according to Allen et al.'s (2017) effective radius estimate presented in Equation 1. This proper length is used only for elliptical, spiral, lenticular, and irregular galaxies with masses around $\log\left(\frac{M_*}{M_\odot}\right) > 10$. These effective radii are doubled to represent an effective diameter used for the semi-major axis length of the ellipses representing these four types of galaxies generated in each bin. Allen et al.'s (2017) equation was found using observations of galaxies between redshifts of $1 < z < 7.2$ with a mass of $\log\left(\frac{M_*}{M_\odot}\right) > 10$. For $z \leq 1$, a value of z = 1 is assumed since galaxy growth has been observed to be higher at high-redshifts, and the full galaxy shape appears to be visually resolved at closer distances in deep field images (Illingworth et al. 2013; Ashby et al. 2015; NASA 2018; De Barros et al. 2019; Allam et al. 2020). This assumption was found to not significantly affect the result and produces an accurate image when using HUDF parameters.

For $z \leq 1$, the corresponding galaxy number densities are multiplied by n_Unseen to account for the total number of galaxies in existence since most low-redshift galaxies are likely detected by the HST. For $z > 1$, if the HST parameters are used in the initial conditions, no further galaxy number densities are adjusted. If the JWST parameters are used in the initial conditions, the rest of the galaxy number densities are multiplied by n_Unseen due to the significantly increased sensitivity of the JWST's NIRCam since z~1 appears to be the point where the number of observed galaxies starts to decrease drastically.

Sixth, for all proper length estimates associated with redshifts of $z > 1$, the r_Increase percentage is multiplied to determine a new proper length since a larger portion of each high-redshift galaxy is likely to be seen due to the increased sensitivity of the JWST compared to HST. Finally, a value of k is chosen based on the value of $\Omega_k$ where k = 0 if $\Omega_k$ = 1, k = 1 if $\Omega_k$ > 1, and k = -1 if $\Omega_k$ < 1. The value of k is used in a future calculation.

### 3.3.3 Extragalactic Background Light

The EBL is calculated for the image based on Cooray's, (2016) average for the optical and near-infrared wavelengths. Cooray (2016) estimates the EBL to be $9.0^{+1.7}_{-0.9} \frac{\text{nW}}{\text{m}^2 \cdot \text{sr}}$, so a value of $9.0 \frac{\text{nW}}{\text{m}^2 \cdot \text{sr}}$ is assumed. A background level of EBL is calculated and incorporated in this simulation as a uniform background noise. EBL is calculated given the intensity, the primary mirror area, the angular dimensions of the detector's field of view, the exposure, and filter wavelength resulting in a value in units of number of photons per pixel. This number is ratioed with the photon saturation level of the respective NIRCam filter to result in an opacity value. A square with this opacity is plot on the deep field image to serve as the first piece on which the deep field image is built.

### 3.3.4 Galaxy Selection

To simulate an ellipse, a galaxy type is first chosen from the "GalaxyPick" function that depends on redshift. This function splits the galaxy types between dwarf-type galaxies and non-dwarf-type galaxies. Dwarf galaxies are known to be the most abundant type of galaxies in the universe, and it is assumed all galaxies began as dwarf galaxies in the early universe (COSMOS - The SAO Encyclopedia of Astronomy n.d.). The ratio of dwarf-type to non-dwarf-type galaxies begins with a value of 1 at z = 0 and evolves proportional to 1 + z. This estimation is made to match the merger rate of Conselice et al. (2016) assuming dwarf galaxies merge into non-dwarf-type galaxies. Once a galaxy is chosen as a dwarf-type galaxy, a designation of dwarf spheroidal or dwarf elliptical is assigned with equal odds at z = 1 and an increasing ratio of dwarf spheroidal to dwarf elliptical proportional to 1 + z. The correct ratio is naturally difficult to determine due to limitations of observations and the faintness of dwarf galaxies, so the above estimations were chosen as a best guess. This also accounts for high-z spirals and elliptical galaxies with only their cores resolved.

Once a galaxy is chosen to be a non-dwarf-type galaxy, it is assigned a type with the following odds: 60% spiral, 10% elliptical, 20% lenticular, and 10% irregular. These odds are selected based on well known estimates of galaxy types (COSMOS - The SAO Encyclopedia of Astronomy n.d.)

Once a galaxy type is chosen, the following galaxy characteristics are assigned: semimajor axis length (kpcs), semiminor axis ratio to semimajor axis length, largest star mass in the galaxy's initial mass function, the total mass of the galaxy, and the total number of stars in the galaxy. No two galaxy types are simulated to be the same, rather the values of these characteristics are randomly assigned within their measured ranges. Table 1 depicts the range of characteristics for each galaxy type derived from Carroll & Ostlie (2018) and COSMOS - The SAO Encyclopedia of Astronomy (n.d.).

The largest star mass is found using the "MassUpperLimitNoStarFormation" function which depends on the age of the universe, the time since star formation first began, and redshift. A calculation of the maximum star mass is found using Equation 11 where $t_{MS}$ is the time since star formation began at the current redshift in years, $t_\odot$ is the estimated lifespan of the sun, and M is the largest stellar mass in the galaxy's IMF (COSMOS - The SAO Encyclopedia of Astronomy n.d.).

$$\frac{t_{MS}}{t_\odot} = \left(\frac{M}{M_\odot}\right)^{-2.5} \quad (11)$$

**Table 1**. Galaxy characteristics are generated in the "GalaxyPick" function that depends on redshift. A galaxy type is chosen, and characteristics assigned using the range of values below. The semimajor axis proper length is calculated from Equation 1 for elliptical, spiral, lenticular, and irregular galaxies. Dwarf galaxies have semi-major axis proper lengths that fall within the ranges listed below. The semi-minor axis is defined by multiplying the semimajor axis proper length by a random number between 0 and 1 if the galaxy is not a dwarf spheroidal. Dwarf spheroidal galaxies have semiminor axis proper lengths equal to their semimajor axis. The total mass of each galaxy is selected from within their observed mass ranges divided by 1 + z due to the merger rate measured by Conselice et al. (2016). The total stellar mass in each galaxy is calculated by multiplying the total mass by $\frac{1}{5}$ as approximately $\frac{1}{5}$ of a galaxy's mass is believed to be dark matter (NASA 2022). Another factor is multiplied for every galaxy except for lenticular galaxies due the interstellar medium's (ISM) mass. There is a wide range of possible masses for a galaxy's ISM and no way to accurately predict an ISM based on galaxy type due to the wide range of processes that may alter the ISM mass. Estimates range anywhere from 1-50% of a galaxy's stellar mass as estimated by University of Maryland (n.d.). A random ISM percentage is calculated for each galaxy aside from lenticular galaxies and divided from the baryonic mass of each galaxy. The total number of stars of each galaxy is calculated by dividing the total mass by an average star mass of 0.3 $M_\odot$ to estimate the number of stars in each galaxy. An average mass of 0.3 $M_\odot$ is based on Kroupa's (2001) estimation. The largest stellar mass in a galaxy's IMF is either selected as 100 $M_\odot$ as is typical for the zero-age main sequence initial mass function (Carroll & Ostlie 2018). For elliptical galaxies and lenticular galaxies, the largest stellar mass is calculated using the "MassUpperLimitNoStarFormation." Lenticular galaxies are assumed to have lost their star-forming gas shortly after formation and elliptical galaxies are assumed to have IMFs ranging somewhere between lenticular and spiral galaxies.

| Galaxy Type | Semi Major Axis Length | Semi Minor To Semi Major Axis Ratio | Total Mass Range ($M_\odot$) | Total Number of Stars | Largest Stellar Mass ($M_\odot$) |
|---|---|---|---|---|---|
| Dwarf Elliptical | 1-10 kpcs | 0-1 | $\frac{1e7}{1+z} - \frac{1e9}{1+z}$ | $M_{total} \times \frac{1}{0.3} \times \frac{1}{5} \times \frac{1}{1+ISM}$ | 100 |
| Dwarf Spheroidal | 0.1-0.5 kpcs | 1 | $\frac{1e7}{1+z} - \frac{1e8}{1+z}$ | $M_{total} \times \frac{1}{0.3} \times \frac{1}{5} \times \frac{1}{1+ISM}$ | 100 |
| Elliptical | Equation 1 | 0-1 | $\frac{1e8}{1+z} - \frac{1e14}{1+z}$ | $M_{total} \times \frac{1}{0.3} \times \frac{1}{5} \times \frac{1}{1+ISM}$ | MassUpperLimit… function − 100 |
| Spiral | Equation 1 | 0-1 | $\frac{1e9}{1+z} - \frac{1e12}{1+z}$ | $M_{total} \times \frac{1}{0.3} \times \frac{1}{5} \times \frac{1}{1+ISM}$ | 100 |
| Lenticular | Equation 1 | 0-1 | $\frac{1e8}{1+z} - \frac{1e14}{1+z}$ | $M_{total} \times \frac{1}{0.3} \times \frac{1}{5}$ | MassUpperLimit… function |
| Irregular | Equation 1 | 0-1 | $\frac{1e8}{1+z} - \frac{1e10}{1+z}$ | $M_{total} \times \frac{1}{0.3} \times \frac{1}{5} \times \frac{1}{1+ISM}$ | 100 |

### 3.3.5 Initial Mass Function

Once a galaxy's characteristics are defined, an IMF is constructed from the galaxy characteristics in Table 1. All IMF's are assumed to follow the same slopes estimated by Kroupa (2001) normalized to an approximately consistent function with no jumps or breaks. The characteristics of the IMF are first defined using the "GalaxyStellarDistribution" function which depends on the number of stars in the respective galaxy and the largest main sequence stellar mass of the galaxy. A list of stellar masses is created with masses from 0.01 $M_\odot$ to the largest main sequence stellar mass in the galaxy in intervals of 0.01 $M_\odot$. For each mass, the IMF is constructed following Kroupa's (2001) estimation in 0.01 $M_\odot$ bins and normalized so integration over the entire IMF results in the galaxy's number of main sequence stars.

Three more lists are created corresponding to the IMF's mass list. The "starT" list holds the surface temperature of a star at each mass in the mass list calculated from Equation 12, the "starNum" list holds the number of stars at each mass in the mass list, and the "starRadius" holds the radius of each star at each mass in the mass list calculated from Equation 13. When these lists are created, "starT", "starNum", and "starRadius" have the same number of elements as the mass list. These three lists are consolidated into a larger list named "GalaxyParameters" and set as the output of the "GalaxyStellarDistribution" function.

$$T = 5778K \times \left(\frac{M}{M_\odot}\right)^{\frac{5}{8}} \quad (12)$$

$$r = \begin{cases} 6.96 \times 10^8 \times \left(\frac{M}{M_\odot}\right)^{\frac{3}{7}} & \text{for } M < 3\ M_\odot \\ 6.96 \times 10^8 \times \left(\frac{M}{M_\odot}\right)^{\frac{15}{19}} & \text{for } M \geq 3\ M_\odot \end{cases} \quad (13)$$

In equations 12 and 13, T is the star's temperature, M is the stellar main sequence mass, $M_\odot$ is the sun's mass, and r is the stellar radius. These equations are derived from homology relations.

Once the IMF of a galaxy is calculated, the galaxy's black body spectrum is considered. The "BlackBodyGalaxyPhotons" function calculates the number of photons received by NIRCam from the galaxy. This function depends on filter wavelength, the previously calculated temperature, stellar number, and radius lists from the "GalaxyStellarDistribution" function, the redshift, the area of the primary mirror, and the period of exposure. The filter's wavelength is input into the function and blue shifted to the rest-wavelength emitted by the galaxy. For each star type in the three lists, the black body radiance is calculated for the rest wavelength and converted to photons. This photon number is multiplied by the corresponding number of stars. This process continues for each element in the three lists until a total number of photons from the galaxy over the exposure period is calculated. This process occurs if the rest-wavelength is $\geq 9.12 \times 10^{-8}$, otherwise the number of photons is set to 0 due to the Lyman break (Steidel et al. 1996). Additionally, the number of photons is halved to simulate approximately half of a galaxy's light being blocked by dust (Driver et al., 2008). Finally, the number of photons is multiplied by $\frac{1}{1+z}$ to account for the decrease in photon flux according to the luminosity distance (Weinberg 1972; Weedman 1988; Sahni & Starobinsky 2000; Balakrishna Subramani et al. 2019; Peebles 2020).

Example blackbody radiation curves of an elliptical galaxy with $300 \times 10^9$ stars and a maximum stellar mass of 15 $M_\odot$, can be seen in Fig. 1. An example lenticular galaxy IMF can be seen in Fig. 2 at a range of redshifts with no star formation. An example spiral galaxy IMF can be seen in Fig. 3 at a range of redshifts and ongoing star formation.

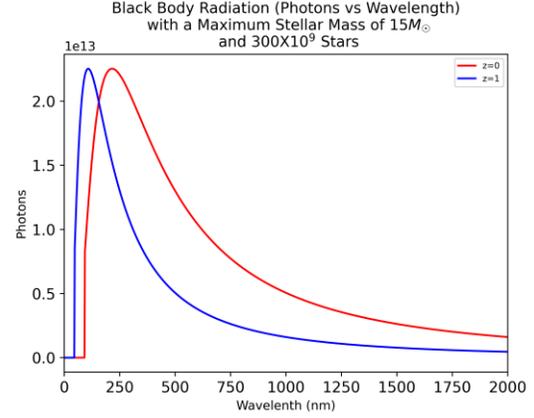

**Figure 1.** An example black body spectrum produced by the "BlackBodyGalaxyPhotons" and "GalaxyStellarDistribution" functions. The blue line represents the spectrum emitted from the galaxy at z = 1, and the redline is the observed spectrum at z = 0. The Lyman Break can be seen by a sudden drop at 91.2 nm

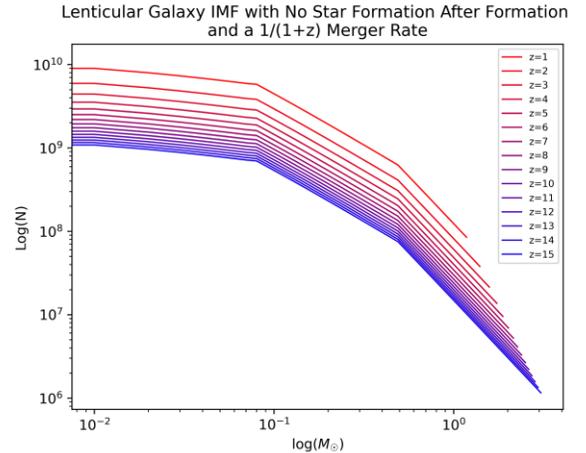

**Figure 2.** An example lenticular galaxy's IMF at multiple redshifts with no star formation occurring since formation. The IMF's upper mass limit and number of stars are both dependent on redshift.

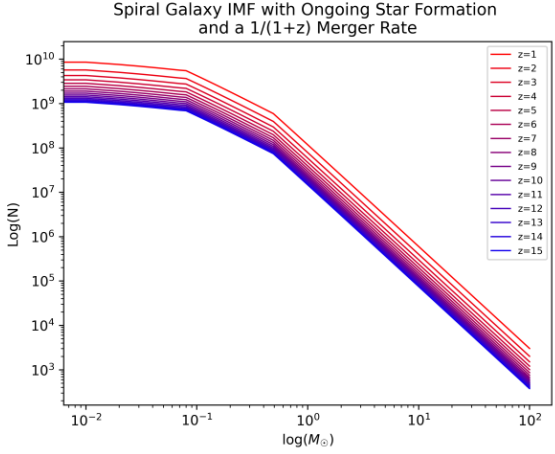

**Figure 3.** An example spiral galaxy's IMF at multiple redshifts with ongoing star formation. Only the IMF's number of stars is dependent on redshift. The IMF lines are closer together due to a consistent star number ratio in each stellar mass bin.

### 3.3.6 Alpha Calculation

Once the number of photons received by NIRCam from a galaxy is calculated, an alpha value is calculated for the ellipse. This alpha value is synonymous with opacity in the matplotlib ellipse module. The NIRCam's bright source limit is converted into number of photons from the listed value for the respective filter designated in the initial conditions using STSci's (2016) values. This conversion from Vega K-band magnitudes for a solar type G2V star to number of photons using the "VegaPhotons21_4seconds" function. The total number of photons received by the detector is divided by the number of pixels occupied by the ellipse to result in number of photons per pixel. This value is divided by the NIRCam's bright source limit for the respective filter to result in an alpha value between 0 and 1 with 1 being fully opaque and zero being transparent. Values >1 represent a saturated source and are set equal to one.

The process described in sections 3.3.4 to 3.3.6 are carried out for every galaxy in each redshift bin until a deep field image is constructed. The number of galaxies of each galaxy type is calculated and output in the command line.

### 3.3.7 Gaussian Filter

The edge of a galaxy is difficult to define when measuring its length and width due to star density generally decreasing as radius increases. The brightness of a galaxy near its edge can be defined De Vaucouleurs' Law and can be seen in Equation 14 where $I(r)$ represents the intensity at radius $r$, $r_e$ is the effective radius, n is the Sérsic index and $I_0$ is the brightness at $r_e$ (de Vaucouleurs 1948; Mazure & Capelato 2002).

$$I(r) = I_0 e^{-7.669\left(\left(\frac{r}{r_e}\right)^{\frac{1}{n}} - 1\right)} \qquad (14)$$

Due to computational constraints, a Sérsic profile could not be applied to each simulated galaxy and are not considered. As a proxy, a gaussian filter was applied to the final plot to smooth out the edges of the small sources. The HUDF was converted into a monochrome negative using an open-source program as a reference to calibrate the gaussian filter (Imgonline n.d.). The HUDF's galaxy coverage percentage was calculated with this simulation's software and found to be 5.7% with a threshold of ≥10% opacity as the baseline comparison. The HUDF was simulated and gaussian filter parameters adjusted to result in a value approximately equal to this coverage percentage. These gaussian filter parameters were then used for all deep field simulations.

### 3.3.8 First and Second Figure Generation

This simulation first creates an image plotting angular diameter vs redshift for an object with a proper length equal to the test length out to the test redshift specified in the control panel. This image is used to test and display the angular-diameter-redshift relation used in this simulation according to Equation 15 where $\theta$ is the angular diameter, $D$ is the proper length, $c$ is the speed of light, and $S$ is defined in Equation 16 (Weinberg 1972; Weedman 1988; Sahni & Starobinsky 2000; Balakrishna Subramani et al. 2019; Peebles 2020).

$$\theta = \begin{cases} \frac{DH_0(1+z)}{cS} & \text{for } k = 1 \\ \frac{DH_0(1+z)\sqrt{|\Omega_\Lambda + \Omega_m - 1|}}{c \sin(S\sqrt{|\Omega_\Lambda + \Omega_m - 1|})} & \text{for } k = 0 \\ \frac{DH_0(1+z)\sqrt{|\Omega_\Lambda + \Omega_m - 1|}}{c \sinh(S\sqrt{|\Omega_\Lambda + \Omega_m - 1|})} & \text{for } k = -1 \end{cases} \quad (15)$$

$$S = \int_0^z \frac{dz'}{\sqrt{\Omega_m(1+z')^3 + (1-\Omega_m-\Omega_\Lambda)(1+z')^2 + \Omega_\Lambda(1+z')^m}} \quad (16)$$

A second image is generated in the same manner using the basic trigonometry in Equation 17 instead of Equations 15 and 16. This image represents the expected angular diameters if the universe was not expanding and is used to compare the effects of expansion on the apparent angular diameter of high-redshift objects.

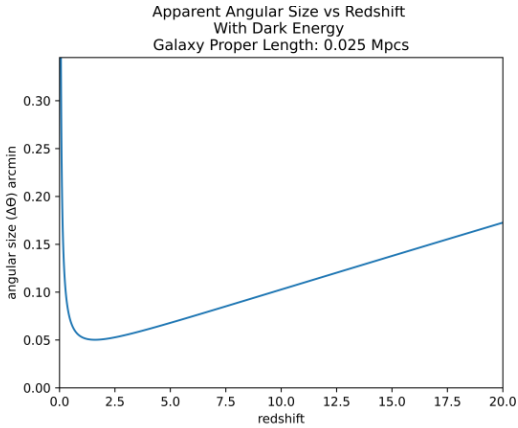

**Figure 4.** An example of the first image generated by this simulation. This image displays the expected change in apparent angular diameter at distances associated with redshift due to the expansion of the universe.

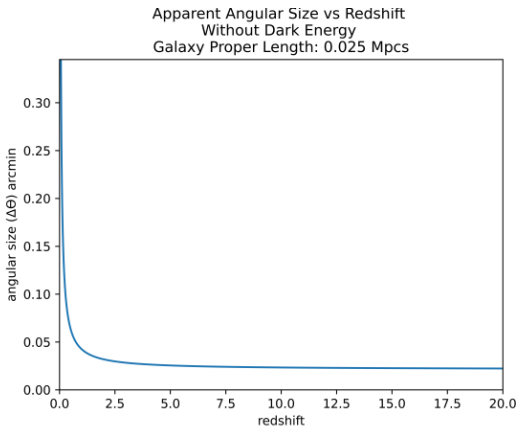

**Figure 5.** An example of the second image generated by this simulation. This image displays the expected change in apparent angular diameter at distances associated with redshift if the universe was not expanding. This image is used for reference in conjunction with the second generated image for a visual reference to the expected angular diameter changes from the universe's expansion.

$$\theta = \arctan\left(\frac{\text{Proper Length}}{\text{Proper Distance}}\right) \quad (17)$$

Figs. 4 and 5 present examples of the first and second images generated by this simulation respectively. This comparison is useful to visually see the difference between the difference between the expected angular size of a high-z galaxy and its apparent angular size based on the selected mass and energy densities. At high redshifts, galaxies appear to be several times larger than expected.

### 3.3.9 Third Figure: The Deep Field Simulation

The third figure displays a visual representation of the deep field simulation. The code runs through each galaxy number density bin from the highest redshift bin to the lowest redshift bin and generates ellipses on a two-dimensional plane for each bin using the processes described in sections 3.3.4 to 3.3.6. Since the galaxy number density estimates describe the number of galaxies per arcmin$^2$, the number of galaxies is multiplied by the total area in arcmin$^2$ according to the specifications in the control panel. An ellipse with a random spatial and angular orientation is generated for each galaxy with a semi-major axis and semi minor axis defined by the galaxy type. Each ellipse's angular orientation, semi-major axis, and semi-minor axis are adjusted according to the effects of dark energy anisotropy that will be discussed in section 3.3.11. After a plane of ellipses is generated for each bin, the planes are added together to form one deep field image and displayed as the third figure. An example of the third generated figure can be seen in Fig. 6.

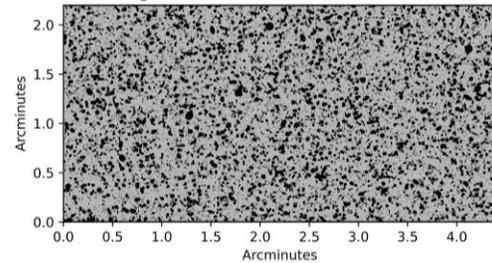

**Figure 6.** An example of the third image generated by this simulation. This image displays a visual depiction of the JWST deep field simulation with randomly generated positions and orientations of galaxies and systematically generated galaxy types represented by ellipses. This image assumed 76959 galaxies are observable by the JWST with a furthest measurable redshift of z = 15.

### 3.3.10 Fourth and Fifth Figures: Reference Figures

The fourth and fifth figures are generated to produce an image with a 100% galaxy coverage percentage and a 0% galaxy coverage percentage respectively. These figures do not represent a deep field simulation of any kind and are only used to calculate the galaxy number density of the third figure.

The creation of these reference figures are vital to the success of the galaxy coverage percentage calculation. The number of pixels with an alpha value above the EBL in each reference figure is calculated using the cv2 module. The number of pixels in the empty image is subtracted from the full image to determine the total number of pixels in the image's field of view. The deep field image's number of pixels with an alpha value above the EBL is calculated in the same manner and subtracted from the empty deep field image to determine how many pixels are not covered by ellipses. When this value is divided by the total number of pixels in the field of view and multiplied by 100%, a galaxy coverage percentage is found between 0% and 100%. This value is displayed in python's console after the code completes execution. Since these figures simply produce solid black and white squares respectively, examples are not shown in this paper.

### 3.3.11 Sixth Figure: Anisotropy of Dark Energy

As discussed above, the anisotropy of dark energy is defined in the initial conditions and perturbed in the ensemble. This study introduces the following novel effect from dark energy anisotropy by assuming the anisotropy of dark energy can be generalized as a direction-dependent Hubble constant representing a universe expanding at a different rate along one axis than the others.

The angular diameters distance represents the distance of a galaxy when its light was first emitted. This distance is used to calculate the apparent angular diameter of a galaxy as seen today due to the radial expansion of the universe. The increase in size occurs because the galaxy was much closer at the time of emission, and the photon directions are conserved in an expanding universe. An additional source of an angular diameter increases results when considering the expanding space between photons originating from opposite sides of a galaxy. Since direction is conserved, photons require an angle pointed further towards the galaxy's center to overcome the added expansion. This corresponds to a small increase in the apparent angular size of an object. It is worth noting, this effect had a miniscule effect on the results of this simulation.

This effect was accounted for by applying Equations 15 and 16 with a different $H_0$ in the x-axis than in the y-axis. Fig. 7 presents the expected change to a high-z galaxy's appearance where the gray shaded ellipse represents the appearance in a non-expanding universe, the blue ellipse represents the apparent size of the same galaxy caused by isotropic universe expansion, and the red ellipse represents an anisotropic universe expansion where $H_0$ in the y-axis is less than $H_0$ in the x-axis.

The change in a high-redshift galaxy's expected shape takes the form of not just a smaller image but a decreased angle of orientation in the direction of the axis of larger $H_0$. While this effect would be indiscernible in the image of a single galaxy, with a large enough sample of galaxies, an angle bias may appear towards the axis of larger $H_0$.

Once an ellipse is randomly generated in the deep field image, its semi-major axis, semi-minor axis, and angle of orientation are adjusted for the dark energy anisotropy ratio defined in the initial conditions using trigonometry. These adjustments can be seen in Equations 18-20 where $a_{old}$ is the apparent length of the semi-major axis in arcmin of a high-z galaxy in a non-expanding universe, $b_{old}$ is the length of the semi-minor axis in arcmin in a non-expanding universe, $H_{0,y}$ is the Hubble constant in the y-axis, $H_{0,x}$ is the Hubble constant in the x-axis, $a$ is the apparent length of the semi-major axis in arcmin in an expanding universe, $b$ is the apparent length of the semi-minor axis in arcmin in an expanding universe, $\theta_{new}$ is the galaxy's new angle with respect to the x-axis in a universe with anisotropic expansion, and $\theta$ is the galaxy's angle with respect to the x-axis in a universe with isotropic expansion.

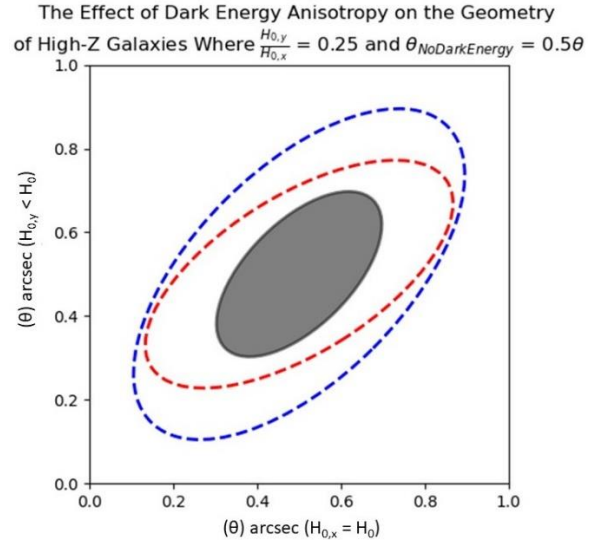

**Figure 7**. A high-z galaxy with a semi-major axis of 0.5 arcmin is represented by 3 ellipses. The gray shaded ellipse represents the galaxy's appearance in a non-expanding universe, the blue ellipse represents the increase in apparent angular size due to the Angular-Diameter-Redshift relation from isotropic expansion of dark energy, and the red ellipse represents the distortion from anisotropic universe expansion. In this image, the expansion of the universe, and the value of $H_0$, is greater in the x-axis than in the y-axis where $H_{0,x}$ and $H_{0,y}$ represents the expansion rates of the universe in the x-axis and y-axis respectively, and where $H_{0,y} = 0.25 H_{0,x}$.

$$\theta_{new} = \arctan\left(\frac{2a_{old}\sin(\theta)+\left(\frac{H_{0,y}}{H_{0,x}}\right)(2a\sin(\theta)-2a_{old}\sin(\theta))}{2a\cos(\theta)}\right) \quad (18)$$

$$2a = \frac{\left(\frac{H_{0,y}}{H_{0,x}}\right)\cos(\theta)}{\cos(\theta_{new})} \quad (19)$$

$$2b = \frac{\left(\frac{H_{0,y}}{H_{0,x}}\right)\sin(\theta_{new})}{\sin(\theta)} \quad (20)$$

Equations 18-20 are valid for all $\theta$'s between 0 and 90 except for $\theta = 90$ and $\theta = 0$ where the limit must be taken to yield the correct result. Equations 18-20 are applied to each simulated galaxy and creates a new ellipse based on its redshift and orientation. These new ellipses are used for the final plot. Equations 18 is adjusted when $\theta > 90$ to ensure $\theta_{new}$ is always closer to the x-axis than $\theta$. This adjustment can be seen in Equations 21-23.

if $\theta \leq 180$, then $\theta = 180 - \theta$ and $\theta_{new} = 180 - \theta_{new}$ (21)

if $\theta \leq 270$, then $\theta = \theta - 180$ and $\theta_{new} = 180 + \theta_{new}$ (22)

if $\theta \leq 360$, then $\theta = 360 - \theta$ and $\theta_{new} = 360 - \theta_{new}$ (23)

Every $\theta_{new}$ is saved in a list used in the final figure. Each $\theta_{new}$ in this list is divided into two new lists depending on $\theta$ being a positive slope or negative slope. All three lists are individually averaged into the average positively oriented ellipses, negatively oriented ellipses, and total ellipses. The average positively and negatively oriented angles are plotted as vectors with a length of 2 arbitrary units in the sixth figure. The average of all angles is displayed in the command line. Isotropic universe expansion yields an average angle of 45 degrees for positively oriented ellipses and 135 degrees for negatively oriented ellipses. An anisotropic universe expansion yields an average angle of < 45 degrees for positively oriented ellipses and > 135 degrees for negatively oriented ellipses. An example of the sixth figure produced by this simulation is shown in Fig. 8. Simulations resulting from Equations 18-20 with varying values of $H_{0,x}$, $H_{0,y}$, and $\theta$ with $\frac{b_{old}}{a_{old}} = 0.5$ and $b_{old} = 0.5$ arcmin are shown in Fig. 9 with the same visual convention used in Fig. 7.

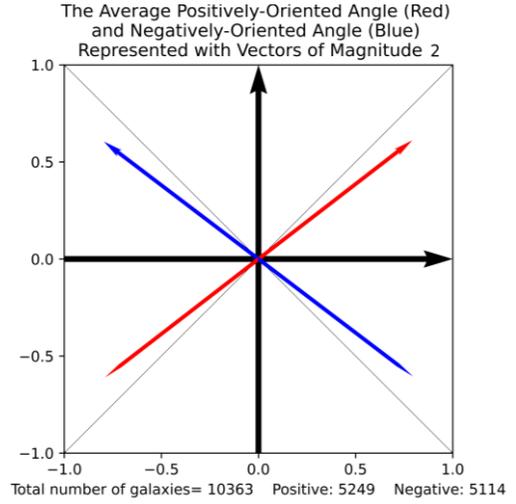

**Figure 8**. An example of the sixth figure produced by this simulation. This figure was produced with DE_Ratio = 0.5. The red arrow represents the average angle of orientation of all positively oriented ellipses and the blue arrow represents the angle of orientation of all negatively oriented galaxies. The red and blue arrows are <45 degrees and >135 degrees due to the anisotropy of dark energy defined in this instance.

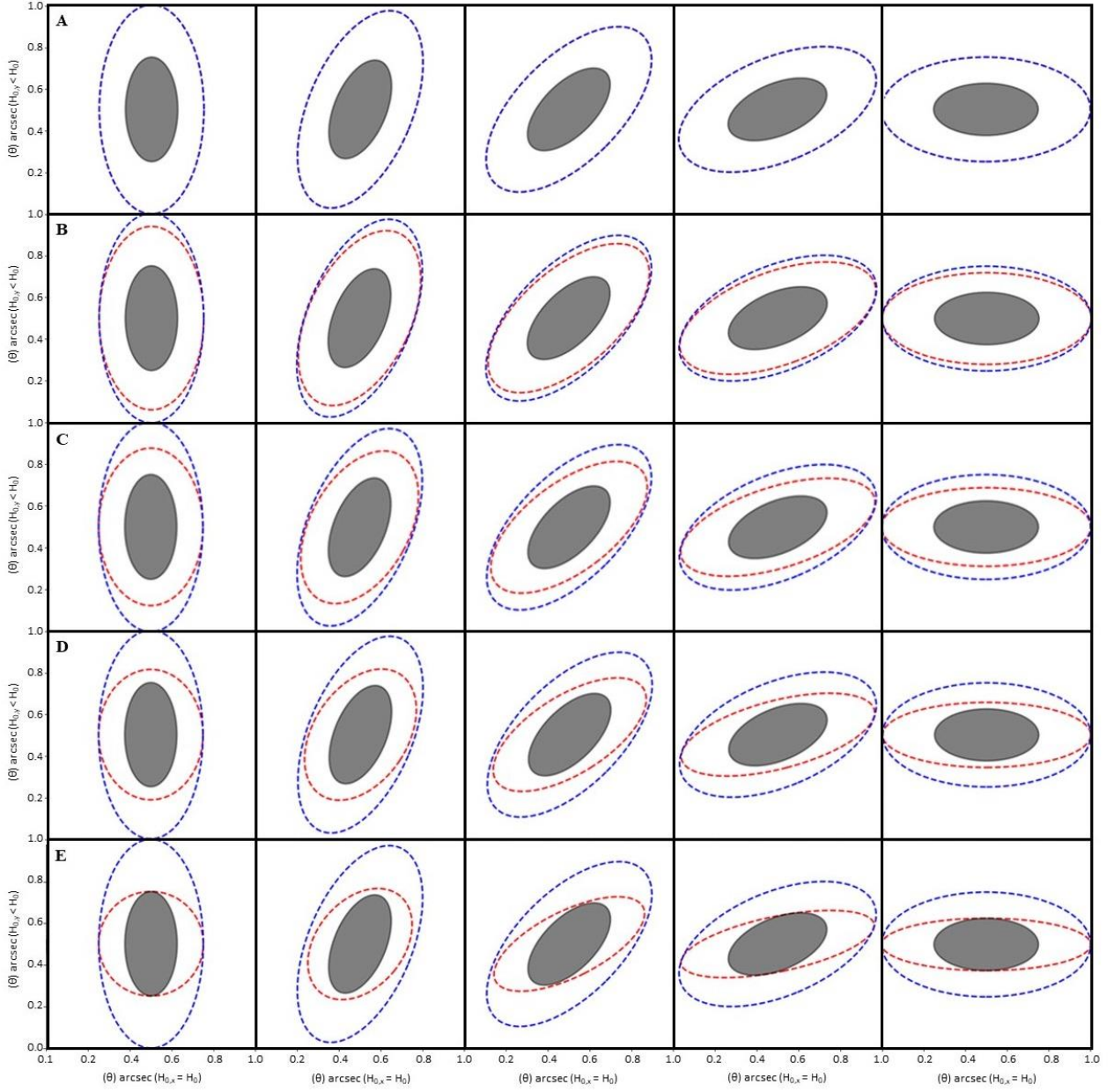

**Figure 9**. Simulations resulting from Equations 18-20 with varying values of $H_{0,x}$ $H_{0,y}$, and θ with $\frac{b_{old}}{a_{old}} = 0.5$ and $b_{old} = 0.5$ arcmin. Rows A, B, C, D, and E represent a simulated high-z galaxy at various angular orientations and DE_Ratios. In these figures, the gray ellipse represents the galaxy's angular size and shape if it was seen at the same distance in a non-expanding universe. The blue dotted ellipse represents the galaxy's apparent angular size due to isotropic expansion of the universe according to the angular-diameter-redshift relation described by Equations 15 and 16. The red dotted ellipse represents the galaxy's new apparent angular size, shape, and angular orientation if the universe was undergoing an overall spatially anisotropic expansion where the expansion rate differs between the x-axis and y-axis. In this figure, the ellipses are at an angle $θ$ between the semi-major axis and the x-axis. Row A represents a high-z galaxy where $H_{0,y} = H_{0,x}$ and $θ$ = 90, 67.5, 45, 22.5, and 0 from left to right respectively. Row B represents a high-z galaxy where $H_{0,y} = 0.75H_{0,x}$ and $θ$ = 90, 67.5, 45, 22.5, and 0 from left to right respectively. Row C represents a high-z galaxy where $H_{0,y} = 0.5H_{0,x}$ and $θ$ = 90, 67.5, 45, 22.5, and 0 from left to right respectively. Row D represents a high-z galaxy where $H_{0,y} = 0.25H_{0,x}$ and $θ$ = 90, 67.5, 45, 22.5, and 0 from left to right respectively. Row E represents a high-z galaxy where $H_{0,y} = 0$ and $θ$ = 90, 67.5, 45, 22.5, and 0 from left to right respectively.

## 4. Results

### 4.1 Code Verification with a HUDF Simulation

This study used a novel approach to simulate deep field images instead of traditional interpolation from observations. While observations were still the basis for simulation, the interpolation to high-redshifts was different from previous studies. Because of this, a test image was created to simulate the HUDF and test the validity of the simulation approach used.

The following initial conditions were chosen to produce a simulation of the HUDF image: jwst_Filter = 'F430M', ageOfStarBirth = 1.75e8, $H_0$ = 71.4, $\Omega_m$ = 0.301, $\Omega_\Lambda$ = 0.706, w = -0.994, n_Unseen = 4, r_Increase = 100%, DE_Ratio = 1, zMax = 11. Additionally, to match the HUDF, the image's dimensions were defined to produce an image size of 3.1 × 3.1 arcmin$^2$, the primary mirror area was defined as 4.52 m$^2$, and an exposure of 134880 seconds was used.

For this HUDF image, the n_Unseen parameter was only applied to redshift planes with z < 1 since the unseen galaxies in the HUDF should remain unseen at high redshifts. A threshold of z = 1 was used as a best guess as determining the number of unseen galaxies is an extremely difficult task as can be seen by the large range of estimates (Hille 2016; Conselice et al. 2016; Lauer et al. 2021).

The simulation was run with these parameters and compared to the HUDF. A monochrome negative of the HUDF was generated using the open-source programme imgonline.com (Imgonline n.d.). The results of the simulation and the HUDF negative are shown side-by-side in Fig. 10.

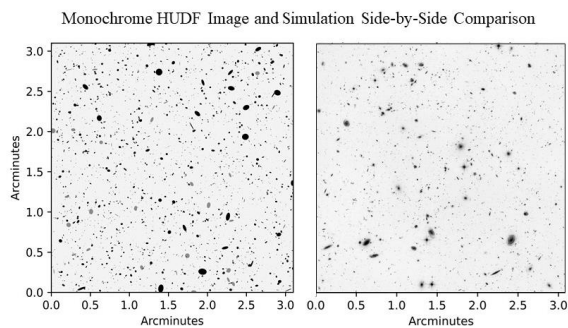

**Figure 10**. A side-by-side comparison of the HUDF Simulation (left) with 10,006 ellipses and a monochrome negative of the HUDF (right) containing an estimated 10,000 galaxies. This has 6.28% coverage percentage compared to 6.9% HUDF.

Several notable differences can be seen between these two images. First, some ellipses in the simulation are uniformly shaded with a distinct edge while galaxies tend to decrease in brightness with an increase in radius. The Gaussian filter smooths out many of the high-z galaxies into to resemble the image of an actual galaxy, but the larger low-z galaxies tend to maintain an edge due to the gaussian filter settings. Second, the spatial distributions, sizes, and brightness of the galaxies differ due to the randomly oriented ellipses with random angular orientation, random spatial placement, systematically generated galaxy types, and random edge-on to top-down orientation for galaxies with elliptical geometries. Third, the EBL alpha is numerically similar with 5.9% opacity in the HUDF and 4.4% in the simulation. While this value differs by a 1.5% interval, the HUDF simulation was run only for the F430M filter at the exposure length of the F435W filter in the HST's Advanced Camera for Surveys (ACS) instrument as a proxy for the HUDF.

Subjectively, the HUDF simulation appears to match well with the HUDF image to include the light grey background contributed from the EBL. Quantitatively, the number of simulated galaxies is consistent with the current estimate of approximately 10,000 and the galaxy coverage percentages of each image above a threshold of 10% opacity is nearly identical with the HUDF having 6.9% compared to 6.28% of the simulation. This can theoretically be fine-tuned further to exactly match the HUDF both in galaxy number, EBL intensity, and galaxy coverage percentage, however, perfecting a replication of the HUDF is beyond the scope of this study. This HUDF replication serves as a validation of the techniques and simplifications used for this simulation.

### 4.2 Parameter Sensitivity Ensemble

A 10-member parameter sensitivity ensemble was run for each parameter as described in section 3.2.1 to determine the most sensitive parameter to changes across its range of uncertainty. Each parameter was perturbed individually across the range of recent observation-based estimates while holding the rest of the initial conditions constant with their respective baseline values. The galaxy coverage percentages were averaged for each parameter and the corresponding standard deviation was found. The parameters were ordered by standard deviation from largest to smallest and can be seen in Table 2.

**Table 2**. The model results are presented showing all 10 perturbation ensembles used to determine the relative sensitivity of the initial conditions. The parameters are listed from highest to lowest standard deviation. The "n_Unseen" parameter was found to be most sensitive to small changes due to a large standard deviation while the dark energy equation of state parameter, "ageOfStarBirth," was surprisingly found to be the least sensitive. All parameters were found to have a positive correlation to the resulting changes in the resulting value of galaxy coverage percentage.

| Parameter | Standard Deviation | Correlation |
|---|---|---|
| "n_Unseen" | 14.40% | Positive |
| "zMax" | 7.90% | Positive |
| "r_Increase" | 6.66% | Positive |
| "jwst_Filter" | 5.29% | Varied |
| $\Omega_m$ | 2.40% | Positive |
| $H_0$ | 1.62% | Positive |
| "de_Ratio" | 1.11% | Positive |
| w | 0.79% | Positive |
| $\Omega_\Lambda$ | 0.32% | Positive |
| "ageOfStarBirth" | 0.21% | Positive |

The results of the parameter sensitivity ensemble should not be interpreted as determining the most sensitive parameters to change in general, but rather, the most sensitive to change from the uncertainty range of their measurements. In other words, a 10% change in the dark energy equation of state parameter (w) may result in a smaller impact than a 10% change in the n_Unseen parameter since the value of w has been measured to within a much smaller range of values than the number of unseen galaxies.

The results of the parameter sensitivity ensemble show that the parameter most sensitive to changes within its measured value is the n_Unseen parameter, the number of unseen galaxies in the HUDF, with a standard deviation of ± 14.40% galaxy coverage percentage. The universe's age when star formation first began is the least sensitive with a standard deviation of just ± 0.21% galaxy coverage percentage.

### 4.3 Ensemble Simulation

The 90-member ensemble was run as described in section 3.2 to determine the expected galaxy coverage percentage in the JWST deep field images. The dimensions of the ensemble simulation were selected to produce an image of the same dimensions and pixel size as specified in the JWST documentation, however, both NIRCam images were combined into one rectangular image (STSci, 2016). The initial condition perturbation ranges are shown in Table 3. The maximum r_Increase value was found by solving equation 14 with n = 1 when the I is increased by a factor of 100 to represent the minimum increase due to the higher sensitivity.

**Table 3**. The perturbation range used for each initial condition in the ensemble. For each parameter's perturbation range, a list was generated of 10 equally spaced values spanning the range. This was done for each of the 9 filters JWST is expected to use for its deep field image resulting in a 90-member ensemble.

| Parameter | Minimum | Maximum |
|---|---|---|
| "n_Unseen" | 2.00 | 10.0 |
| "jwst_Filter" | F090W | F444M |
| "r_Increase" | 100 | 160 |
| "zMax" | 12.0 | 17.0 |
| $\Omega_m$ | 0.2586 | 0.370 |
| $H_0$ | 66.9 | 75.12 |
| "de_Ratio" | 0.850 | 1.00 |
| w | -1.033 | -0.831 |
| $\Omega_\Lambda$ | 0.677 | 0.744 |
| "ageOfStarBirth" | $1 \times 10^8$ | $2.5 \times 10^8$ |

A list of 10 values evenly spaced between the range used for each initial condition was generated. Each set of initial condition values was run 9 times, once for each filter, to produce a 90-member ensemble. Fig. 11 presents a sample of the results from 10 ensemble members in the F277W filter sorted from low to high initial condition values. In Fig. 11, it is evident that a wide variety of results is obtained from the initial condition perturbations, and the confusion limit is not yet reached as resolved sources are not blending into the EBL or each other. In other words, this is the result of galaxy number density saturation and not spatial resolution limitations. The AGW effect is quickly evident in the last 5 samples with more than half of the grey background being covered by galaxies. The galaxy number densities in Fig. 11 range from 25,487 to 342,344. The resulting mean galaxy coverage percentage of the full ensemble run was 47.07 ± 31.85%.

A Sample of 10 Ensemble Members

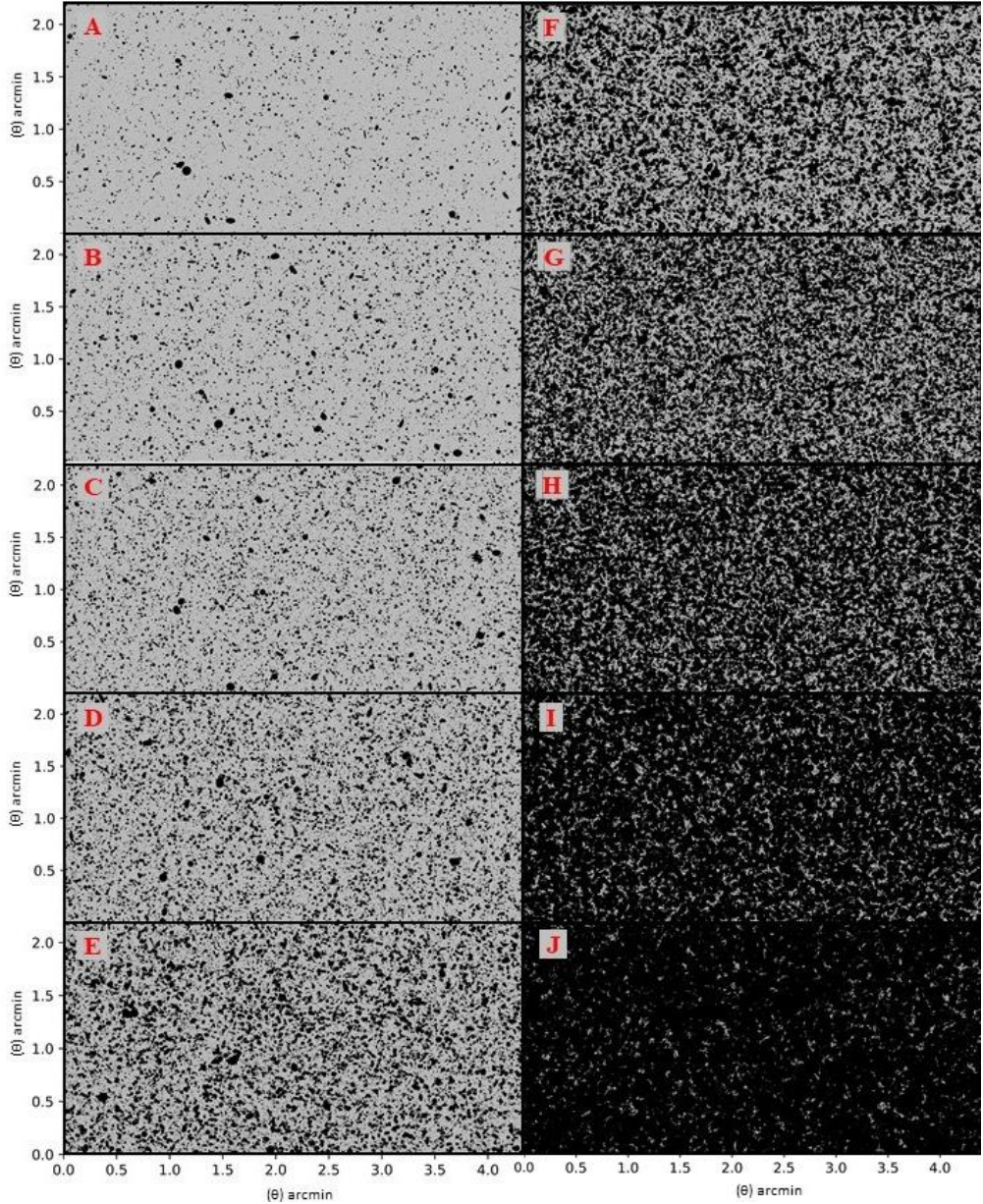

**Figure 11**. The results from 10 ensemble members using the F277W filter. The wide variety of results from the initial condition perturbations can be seen with this ensemble member sample. The ensemble members range from 25,487 to 342,344 simulated ellipses in a deep field image. The high galaxy number densities seen in some of these figures depict the AGW effect. Sections A-J depict the 10 set of initial conditions. Sections A-J depict ten ensemble members from the 90-member ensemble and show how the result changes from the initial condition perturbations.

## *4.4 T-Test Results*

The results from the Ensemble Simulation were analyzed using a one-way one-sample t-test as discussed in section 3.2.3. The t-test failed to reject the null hypothesis under the conditions used in the ensemble, however, a significant portion of the uncertainty range falls within the definition of the AGW effect. Due to the results of the t-test, this study concludes the JWST NIRCam deep field images are not likely to detect the AGW effect from the predicted limitations of the telescope.

## *4.5 Simulation reaching z = 20*

The parameter sensitivity ensemble demonstrated the maximum redshift reached by the telescope had a large associated uncertainty range 7.90% with a maximum redshift of 17 while holding all other initial conditions constant. The final simulation in this range with z = 17 resulted in the furthest galaxies still having a discernable alpha above the EBL indicating galaxies that lie further than the selected redshifts are likely to be seen still. One additional simulation was run keeping all initial conditions constant, but with a maximum redshift of 20 often attributed to the redshift where galaxy formation may have begun (Benson 2010). The resulting galaxy coverage percentage was 61.69% compared with 44.93% at z = 17 with a significant number of high-z galaxies occupying the background of the deep field image.

Increasing the maximum redshift far beyond the predicted range of the JWST still resulted in discernable galaxies, however, this result is obtained by pushing the simulation to its extremes. Admittedly, due to the limitations of current observatories, the characteristics of these galaxies are not well known, so this result is purely theoretical by interpolating known trends and equations backward. Galaxies at these extremes could be fainter, brighter, or simply have not formed yet, so only the high end of redshift theoretically reached by the JWST's NIRCam is considered at z = 17.

*4.6 Monochrome Simulation*

One final simulation was made as a visual prediction of the JWST's deep field image using the F277W filter. The simulation was run again with the following initial conditions: $H_0 = 71.4$, $\Omega_m = 0.301$, $\Omega_\Lambda = 0.706$, w = -0.994, n_Unseen = 2, r_Increase = 130%, DE_Ratio = 1, and zMax = 15. The image dimensions were chosen to match the NIRCam image dimensions including the 44-arcsec gap between the two NIRCam modules (STSci 2016).

The monochrome simulation can be seen in Fig. 12 with 130,349 galaxies and reaching z = 14.22. This monochrome simulation represents this study's official prediction of the JWST deep field image in terms of the galaxy coverage percentage, the EBL, and the middle range of all ensemble initial conditions.

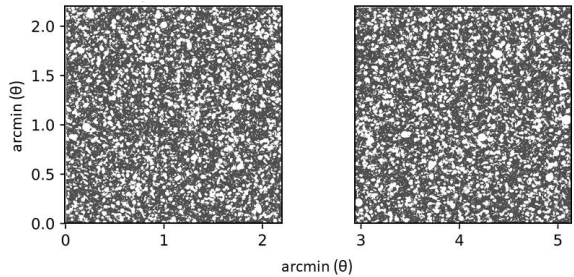

**Figure 12.** A monochrome simulation of this study's JWST deep field images prediction.

## 5. Discussion

The 90-member ensemble simulation resulted in an estimated 47.07% galaxy coverage percentage with an uncertainty of ±31.85%. This percentage was tested for statistical significance above the AGW effect definition of ≥ 50% using a one-way one-sample t-test and found a null result. While the ensemble average was nearly at the AGW definition, the t-test used must be above the 50% threshold for a statistical significance to be less than ~1 even with a large portion of the uncertainty above the threshold.

This is demonstrated further in the parameter sensitivity ensemble where several of the individual parameters have a significant range of estimated values. The majority of previous JWST simulations assumed specific values for $H_0$, $\Omega_m$, and $\Omega_\Lambda$. The parameter sensitivity ensemble demonstrated the variability in results from the uncertainty in these measurements alone. Notably, $\Omega_m$ and zMax alone resulted in substantially larger galaxy coverage percentages. The number of unseen galaxies is unsurprisingly the largest uncertainty by a large magnitude providing a strong case for careful consideration to simulations based on galaxy catalogs alone. This study concludes the assumed universe model has a significant effect on the results of a simulation, and previous simulations may not adequately represent the full range of possible results.

Despite the high statistical significance, this prediction should be understood as a possibility that the JWST will observe the AGW effect if many of the measured parameters lie on the higher end of their uncertainties. This work ultimately concludes the JWST is not likely to observe the AGW effect with greater than 50% of the background being obstructed by galaxies, however, the possibility cannot be ruled out due to the large uncertainty range from the 90-member ensemble residing above the AGW threshold.

Should the AGW effect be observed, a pseudo-cosmological horizon would be found in that an obstruction caused by foreground galaxies may limit the potential of future observatories aiming to reach further distances due to the obstruction. Additionally, this effect may have the potential to adversely affect the funding and scientific goals of larger and more sensitive observatories even if the JWST does not pass the 50% galaxy coverage percentage threshold. The Extremely Large Telescope (ELT) is currently under construction at the time of this research and maybe the first telescope to have trouble observing objects beyond the AGW if the AGW does occur in nature (ESO 2014).

Future work could include accounting for the distortion effects from gravitational lensing when two galaxies at different distances line up. Additionally, the distance to the AGW in terms of redshift could be found through optimization techniques where a 50% galaxy coverage percentage is achieved. Additionally, once the JWST deep field images are released, a comparison can be made to determine the accuracy of all previous simulations. Finally, the approach used in accounting for the effects from dark energy anisotropy may be useful in detecting anisotropy in the normal plane to the line of sight in future deep field images and may be worthy of further investigation in a separate study.

## 6. Conclusion

Throughout this study, it is evident that simulations of deep field images are difficult to confidently predict due to the number of assumptions required and wide ranges in the uncertainty of many constants governing the evolution of the universe. There have been several previous simulations of the JWST's deep field image that used assumptions of the universe's geometry, the galaxy number densities, and the evolutions of high-z galaxies to make their predictions. While these assumptions are built off recent observations, the corresponding wide ranges of uncertainty can lead to a wide range of initial conditions resulting in substantially different results as the parameter sensitivity ensemble demonstrated. A novel geometric-focused deep field simulation of the expected JWST future deep field images was conducted using an ensemble approach. The AGW effect was introduced and defined as $\geq 50\%$ of a deep field image occupied by galaxies with larger than expected apparent angular sizes due to the universe's expansion. A novel effect was introduced resulting from anisotropic expansion of dark energy. A 90-member ensemble was run by perturbing the initial conditions through ranges of uncertainty obtained from recent estimates. A galaxy coverage percentage was calculated for each ensemble member, averaged together, and analysed using a one-way one-sample t-test to determine whether the JWST is likely to observe the AGW effect. This work concluded the JWST is not likely to observe the AGW effect with a galaxy coverage percentage of $47.07 \pm 31.85\%$, but due to the large portion of uncertainty lying above the 50% threshold, this study does not rule out the effect as a possibility. Additionally, the most sensitive parameter to changes within its range of uncertainty was found to be the estimated number of unseen galaxies in the HUDF. The potential impacts of the AGW effect were discussed and its potential to form a pseudo-cosmological horizon that may inhibit the effectiveness of future observatories.

## Acknowledgments

I thank my thesis advisor and colleagues for their indispensable support, encouragement, and expertise in this study. Additionally, I thank the American Military University for allowing me to conduct this study for partial fulfillment of the requirements for the degree of Master of Science in Space Studies. Finally, I thank friends and family for their continued support throughout this endeavor.

## Data Availability

The software written for this study is free and accessible through the Astrophysical Source Code Library (Sailer 2021). This simulation was run on Google Colaboratory and any replication should be run on the same platform. Please report any bugs to the lead author using the contact information listed on the first page of this article.